\documentclass[sigconf, nonacm]{acmart}

\AtBeginDocument{%
  \providecommand\BibTeX{{%
    \normalfont B\kern-0.5em{\scshape i\kern-0.25em b}\kern-0.8em\TeX}}}

\copyrightyear{2024}
\acmYear{2023}
\setcopyright{rightsretained}
\acmConference[arXiv '23]{arXiv}{July 10, 2023}{}
\acmBooktitle{arXiv}
\acmDOI{10.1145/3563657.3596090}
\acmISBN{978-1-4503-9893-0/23/07}

\usepackage{caption}
\usepackage{subcaption}
\usepackage{booktabs}

\begin{document}

\title{ARLang: An Outdoor Augmented Reality Application for Portuguese Vocabulary Learning}

\author{Arthur Caetano}
\email{caetano@ucsb.edu}
\orcid{0000-0003-0207-5471}
\affiliation{
  \institution{University of California, Santa Barbara}
  \city{Santa Barbara}
  \state{CA}
  \country{USA}
  \postcode{93106}
}

\author{Alyssa Lawson}
\email{a_lawson@ucsb.edu}
\orcid{0000-0001-8658-1261}
\affiliation{
  \institution{University of California, Santa Barbara}
  \city{Santa Barbara}
  \state{CA}
  \country{USA}
  \postcode{93106}
}

\author{Yimeng Liu}
\email{yimengliu@ucsb.edu}
\orcid{0000-0002-6742-2908}
\affiliation{
  \institution{University of California, Santa Barbara}
  \city{Santa Barbara}
  \state{CA}
  \country{USA}
  \postcode{93106}
}

\author{Misha Sra}
\email{sra@ucsb.edu}
\orcid{0000-0001-8154-8518}
\affiliation{
  \institution{University of California, Santa Barbara}
  \city{Santa Barbara}
  \state{CA}
  \country{USA}
  \postcode{93106}
}

\renewcommand{\shortauthors}{Caetano, et al.}

\begin{abstract}

With recent computer vision techniques and user-generated content, we can augment the physical world with metadata that describes attributes, such as names, geo-locations, and visual features of physical objects. To assess the benefits of these potentially ubiquitous labels for foreign vocabulary learning, we built a proof-of-concept system that displays bilingual text and sound labels on physical objects outdoors using augmented reality. Established tools for language learning have focused on effective content delivery methods such as books and flashcards. However, recent research and consumer learning tools have begun to focus on how learning can become more mobile, ubiquitous, and desirable. To test whether our system supports vocabulary learning, we conducted a preliminary between-subjects (N=44) study. Our results indicate that participants preferred learning with virtual labels on real-world objects outdoors over learning with flashcards. Our findings motivate further investigation into mobile AR-based learning systems in outdoor settings.

\end{abstract}

\ccsdesc[500]{Human-centered computing~Mixed / augmented reality}

\maketitle

\section{Introduction}
\label{sec:intro}

Learning a second language is a desirable goal for many people worldwide. For native English speakers, it can take anywhere from 600 hours or 24 weeks (French, Spanish) to 2200 hours or 88 weeks (Mandarin, Japanese) of dedicated classroom instruction to reach general proficiency~\cite{zurawsky2006foreign}. Many language learners often do not have the time or resources to dedicate their lives to intensive or immersive instruction. In particular, the primary challenge is to maintain intrinsic motivation for self-directed learning with competing demands on learning time from work, family, and social life \cite{dornyei1998motivation}. 

One solution has been mobile-assisted language learning (MALL), which is popular among second language (L2) learners. Its popularity grew even more during the COVID-19-induced lockdowns when traditional in-person classroom learning was unavailable. For example, Duolingo gained 30 million new users in March 2020 for over 500 million accounts \cite{blanco2020duolingo}. Although popular, previous studies indicate that pedagogical shortcomings, such as the ``reliance on decontextualized grammar-translation exercises'' limit the effectiveness of MALL applications and recommend combining them with more contextualized learning practices \cite{golonka2014technologies, loewen2019mall}. According to the \textit{situated learning theory}, contextualizing learning experiences is essential for improved learning outcomes \cite{lave1991situated, brown1989situated}. While mobile apps have integrated contextually relevant content for L2 learning \cite{al2009approach,beaudin2007context,cui2005context}, the localization is coarse using GPS and content is not spatially embedded. Evidence supports that displaying information in the physical space favors memorization, which makes the spatial context relevant for instructional design \cite{rosello2016nevermind, fujimoto2012memorization}. Thus, in contrast with non-AR mobile apps, mobile AR powered by computer vision tracking is ideally suited to exploit opportunities for ``micro-learning'' \cite{gassler2004integrated} 
grounded in the real world \cite{dunleavy2014augmented}, allowing for both contextualization and spatialization of learning content.

In this paper, we present ARLang, a mobile AR application for visualizing bilingual labels attached to physical objects outdoors to support the micro-learning of language in its spatialized context of use. We based our rationale for contextual and spatialized micro-learning on three findings from cognitive psychology:
\begin{itemize}
\item Situated cognition \cite{brown1989situated} posits that learning that occurs contextually is startlingly fast and successful. The recall is improved when learning and retrieval contexts share perceptual cues \cite{tulving1973encoding}. Using AR outdoors, it is possible to embed knowledge into everyday objects people encounter, providing an opportunity for contextualized and situated language learning. 
\item Spatial memory \cite{yates2013art} and the effectiveness and longevity of the method of ``loci'' suggest a natural human tendency to use spatial context to memorize and recall information. Thus, if L2 content is directly attached to familiar physical objects, it may improve recall. 
\item Motivation \cite{dornyei1998motivation} drives persistence in language learning. Enabling learners to control their learning environment and consider their preferences in instructional design can enhance motivation and engagement \cite{malone2021making,ilin2022role}. Enjoyable engagement is perceived by both teachers and learners as a motivator that encourages participation, improves optimism for learning outcomes, and makes learners more receptive to complex topics \cite{lucardie2014impact,ilin2022role}. 

\end{itemize}

We evaluated ARLang in a between-subject user study for learning Portuguese vocabulary, comparing short-term recall and aspects of user experience, such as preferability, motivation to learn, effort, and enjoyment.

Our results show that learning with ARLang led to a better user experience with a stronger motivation to use ARLang for learning in the future than flashcards.

\section{Related Work}

In this section, we summarize relevant research on AR for second-language learning and mobile learning in general. We highlight theoretical support as well as empirical evidence of the impacts of contextual tools and AR on this domain.

\subsection{Mobile Learning}

Mobile devices support learning across multiple contexts including formal education~\cite{crompton2017use, crompton2018use}, often incorporating gamification elements such as rewards, progress points, and badges~\cite{govender2020survey, shortt2023gamification}. ARLang uses tag interaction counters as a gamification element as seen in Fig.~\ref{fig:slider_and_counter}. Mobile learning has also been used to support self-guided field trips with similar affect and learning outcomes~\cite{shinneman2020self}. ARLang aims to enable self-guided L2 learning outside of the traditional classroom by displaying learning material contextualized in real-world locations using AR.

\subsubsection{Pedagogical Strategies for Mobile Learning}

\textit{Micro-learning} emphasizes learning at the micro-scales of content, time (between $30$ sec and $5$ min), and learner groups \cite{hug2007micro, jahnke2020unpacking}. According to previous research, combining MALL and micro-learning pedagogical strategies boosts learning effectiveness \cite{nikou2018micro, edge2012memreflex}. Examples of mobile micro-learning use cases include learning while walking from home to class or waiting in line \cite{cai2015wait}. \textit{Situated learning} theory states that learning is a social activity immersed in a cultural and physical context \cite{lave1991situated, brown1989situated}. Learning a topic in the actual physical location and cultural context where it naturally manifests can lead to better learning outcomes \cite{dawley2014situated}. Previous work has explored the design of situated AR learning experiences and reported positive learning results \cite{chew2019designing}. The \textit{cognitive theory of multimedia learning} proposes to encode and present learning material in auditory and pictorial forms enabling learners to build mental representations that take advantage of the synergy between stimuli \cite{mayer2022multimedia}. ARLang supports multimedia learning, situated learning, and micro-learning pedagogical approaches by encoding small portions of learning content (vocabulary) as audio and visual elements presented next to objects in a familiar outdoor space.

\subsubsection{Contextual Mobile Learning}
Previous research has explored several solutions to infer spatial context in learning applications. In MicroMandarin \cite{edge2011micromandarin}, researchers used mobile GPS to present context-aware flashcards in a frequency-based strategy. A 4-week field study showed a positive impact on user experience. Earlier work in ubiquitous learning contextualized questions in the learner's surroundings using RFID (radio frequency identification) tags attached to physical objects \cite{ogata2004context, ogataTango, beaudin2007context}. While RFID and GPS provide spatial context, they do not offer the ability to register learning content to a physical location as easily as AR does. Even high-end GPS has an average precision of 2 meters \cite{van2015gps}, and RFID tags require creating and placing them in the study location, which limits their use for ubiquitous learning anytime and anywhere. ARLang uses markerless AR tracking provided by the underlying AR platform to present learning content attached to objects with high precision and without the need to pre-plan or print AR markers.

\subsection{L2 Learning with Mobile AR}

The widespread adoption of mobile devices makes mobile the most popular AR modality, especially given the high cost, low social acceptance and less portable form factor of head-mounted AR devices. Researchers have explored the effects of L2 learning enabled by mobile AR tools, as shown in a meta-review \cite{parmaxi2020augmented}.

\subsubsection{Indoors}
Learning with AR in an indoor environment may not necessarily be transferable to the outdoor setting due to distracting factors and variances in environmental conditions. Researchers have compared a mobile AR app for L2 learning indoors with a non-AR alternative \cite{santos2016augmented} and observed non-significant differences in immediate post-test scores between the two conditions. Conversely, there was a significant difference in user satisfaction in favor of the AR condition. Another research work developed a mobile AR app capable of creating language learning content contextualized to the learner's indoor surroundings \cite{draxler2020augmented}. Despite the context-aware capabilities of the app, it did not improve learners' language skills nor offer improved usability compared to an alternative app that displayed the same learning content as static pictures. Other similar studies have qualitatively analyzed the motivational benefits of using an indoor mobile AR app to learn hiragana calligraphy \cite{yang2018hiragana}. ARLang builds on prior work on using AR for language learning but differs in the setting by embedding content directly into the learner's real-world context outdoors.

\subsubsection{Outdoors}
A substantial research effort has explored geo-localization in mobile devices to support foreign language acquisition in outdoor contexts with AR. An early user study claims that outdoor AR can contextualize foreign language content to outdoor scenic spots and improve foreign language skills \cite{liu2013essays}. To support this claim, authors collected essays written by users (N$=5$) of an outdoor AR English learning app. They evaluated the number of words taught in the app that were present in those essays, but their analysis did not include recall measurements. Regarding user interface design, their application did not register learning content in space but instead used AR elements as shortcuts for a 2D interface that displays the learning material. In another exploration of outdoor mobile AR for L2 learning, researchers found improvements in L2 learning outcomes using Vocabura, a location-based audio application \cite{hautasaari2019vocabura}. In the same publication, authors suggested future work to employ visual AR to increase spatial contiguity between learning material and physical surroundings, which the theory of multimedia learning also proposes \cite{mayer2022multimedia}. ARLang implements spatial contiguity by displaying visual components registered to physical objects (Figure \ref{fig:artag}) in addition to audio content, which is known to be beneficial for learning \cite{mayer2022multimedia}.

\subsection{L2 Learning with Head-mounted AR Displays}

Substantial research shows a positive association between learning and HMD AR devices, as summarized in a meta-review \cite{parmaxi2020augmented}. Unlike mobile AR, HMD AR enables learners to engage with content in hands-free interactions.

\subsubsection{Indoors}
Previous research has shown that learners who used ARBis Pictus, an AR language learning application designed for indoor use, outperformed by 7\% those who used a flashcard web application on productive recall tests administered on the same day \cite{ibrahim2018arbis}. Similarly, VocabulARy \cite{weerasinghe2022vocabulary} evaluated an indoor HMD AR application for Japanese vocabulary learning with keyword visualization compared with a non-AR approach. Their results show better short-term retention for users of the AR condition. These findings quantify the benefits of AR for L2 vocabulary learning and motivate further exploration of the AR learning domain in other environments. The user study we present in this paper explores whether similar benefits occur in an outdoor context or whether the effectiveness of AR for L2 learning varies based on the physical environment. However, our work uses a handheld mobile AR device instead of a head-worn AR device.

\subsubsection{Outdoors}
Prior work has used object detection and machine translation to generate L2 learning content potentially situated outdoors with AR HMDs \cite{vazquez2017serendipitous}. To contribute to the vision of AR language learning outdoors, we provide previously unavailable quantitative results that compare the effectiveness of a mobile AR app and a flashcard app on an L2 vocabulary acquisition task. Recently, researchers have obtained positive results using micro-learning on L2 learning web applications \cite{arakawa2022vocabencounter}. The same publication demonstrates a proof-of-concept of using AR HMDs for L2 vocabulary learning during outdoor strolls but does not report learning performance results of the AR version. Nevermind \cite{rosello2016nevermind}, an outdoor HMD AR memorization app inspired by the ``loci'' method or the memory palace technique, improved participants' short-term recall compared to a paper-based approach, though the task was not related to language learning. Authors claim the higher memorization effect is associated with the fact that both spatial navigation and memorization stimulate the same brain structure, the hippocampus \cite{maguire2003remembering}. Memorization is a necessary cognitive process for learning a new language \cite{khoii2013memorization, yang2011rote}. ARLang builds on the relationship of spatial navigation and memorization to evaluate AR use outdoors in the context of an L2 vocabulary learning task.

\begin{figure*}[!ht]
    \begin{subfigure}[b]{0.35\textwidth}
         \centering
         \includegraphics[width=\textwidth, height=5.5cm]{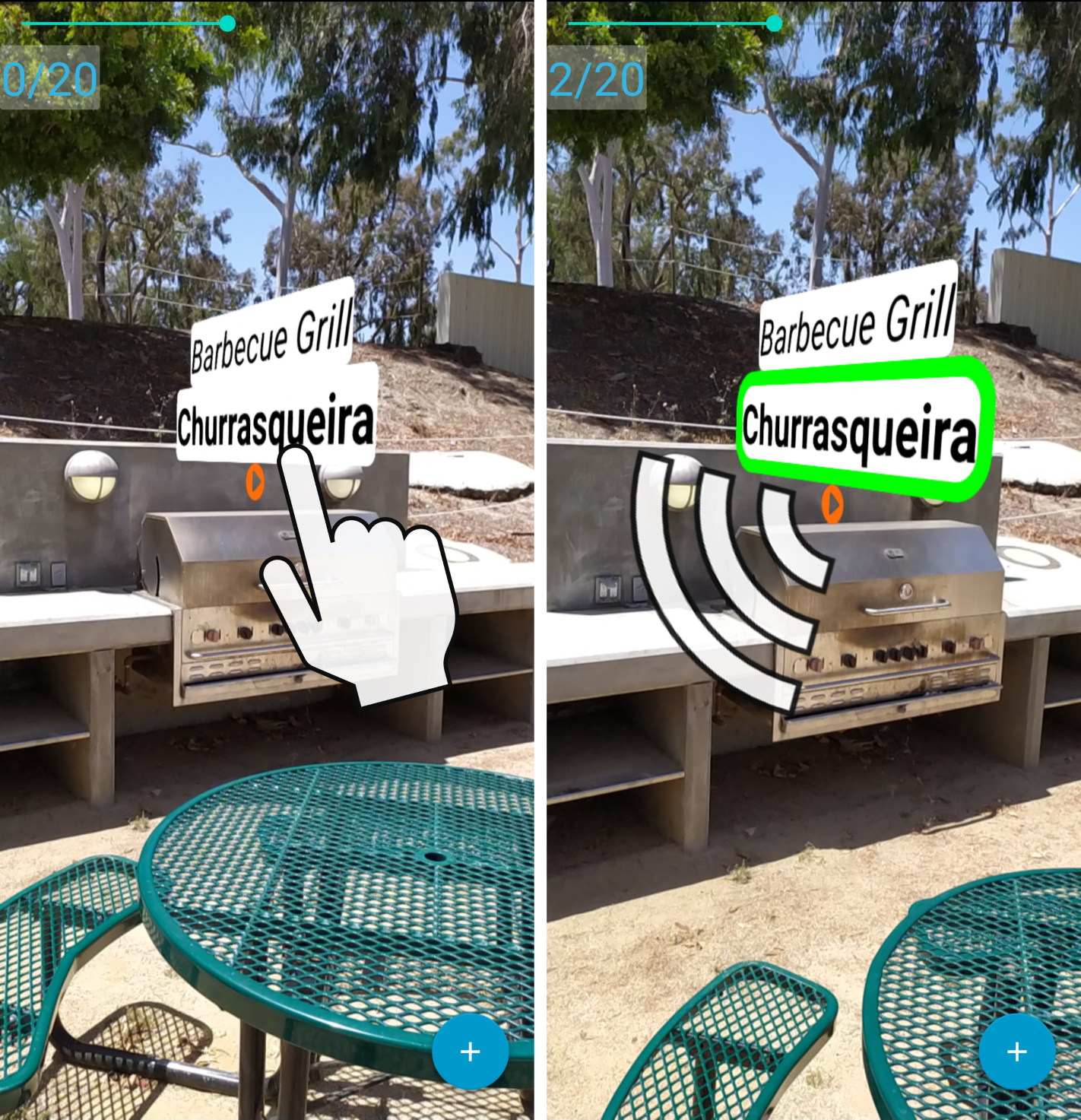}
         \Description[Tapping on an AR tag]{There are two frames. Both depict a barbecue grill as seen through the camera of a mobile device from the same perspective with the UI of the ARLang around the edges. On top of the physical barbecue grill, there is an augmented reality tag with the word “Barbecue Grill” and its Portuguese translation, “Churrasqueira”. In the first frame, there is an icon of a pointing finger on top of the tag denoting that the user can tap the tag. In the second frame, there is the icon of sound waves depicting the sound of pronunciation of “Barbecue Grill” in English and Portuguese. Also in the second frame, the edges of the tag UI element are highlighted in green.}
         \caption{}
         \label{fig:artag}
     \end{subfigure}
     \hspace{0.01\textwidth}
     \begin{subfigure}[b]{0.35\textwidth}
         \centering
         \includegraphics[width=\textwidth, height=5.5cm]{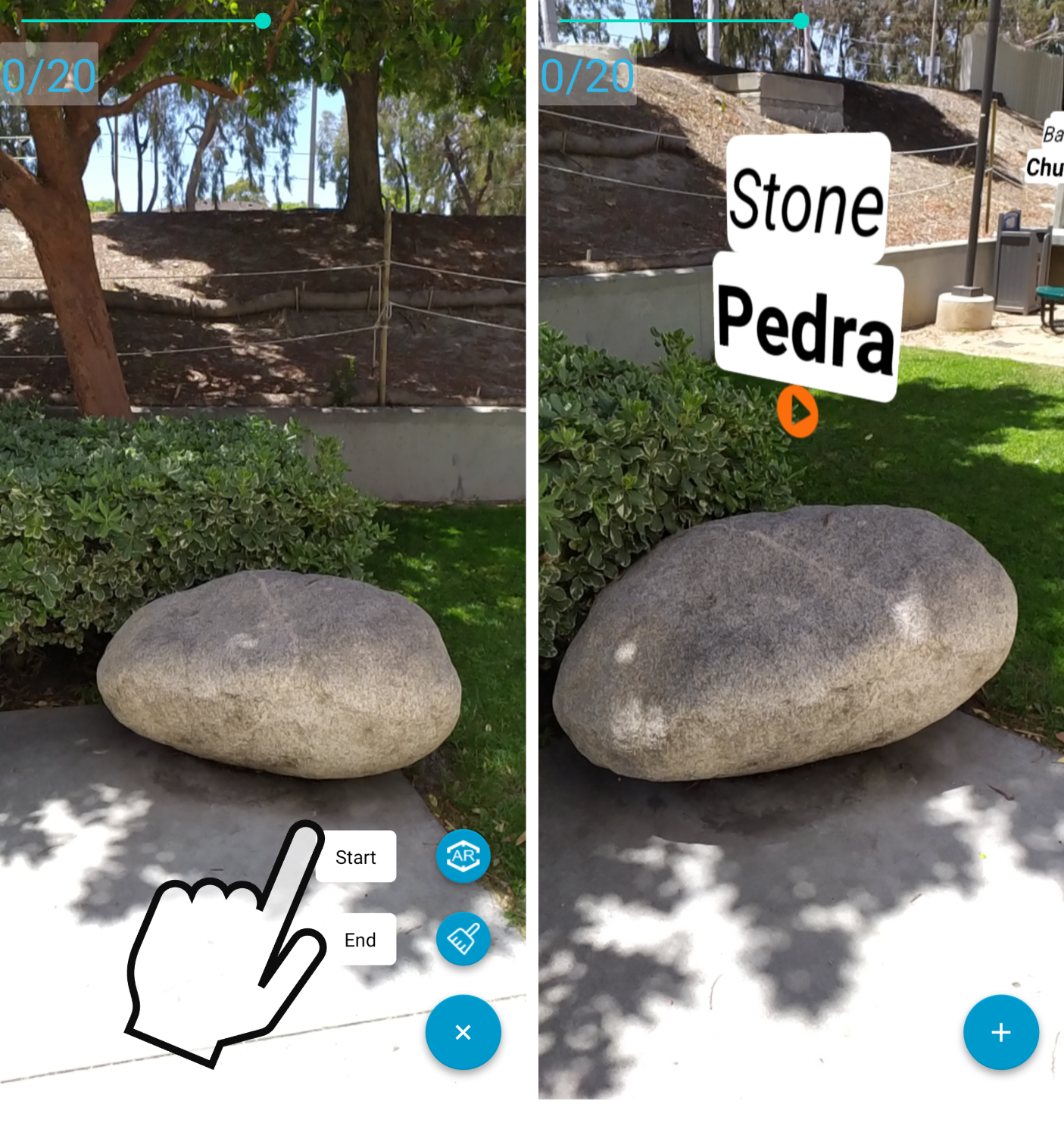}
         \Description[Initializing ARLang]{There are two frames. Both depict a large stone as seen through the camera of a mobile device from the same perspective with the UI of the ARLang around the edges. In the first frame, there is an icon of a pointing finger tapping on a button on the button right corner of the screen that says “Start”. In the second frame, an augmented reality tag with the word “Stone” and its Portuguese translation, “Pedra”, appear on top of the large stone.}
         \caption{}
         \label{fig:foldable}
     \end{subfigure}
     \hspace{0.01\textwidth}
     \begin{subfigure}[b]{0.18\textwidth}
         \centering
         \includegraphics[width=\textwidth, height=5.5cm]{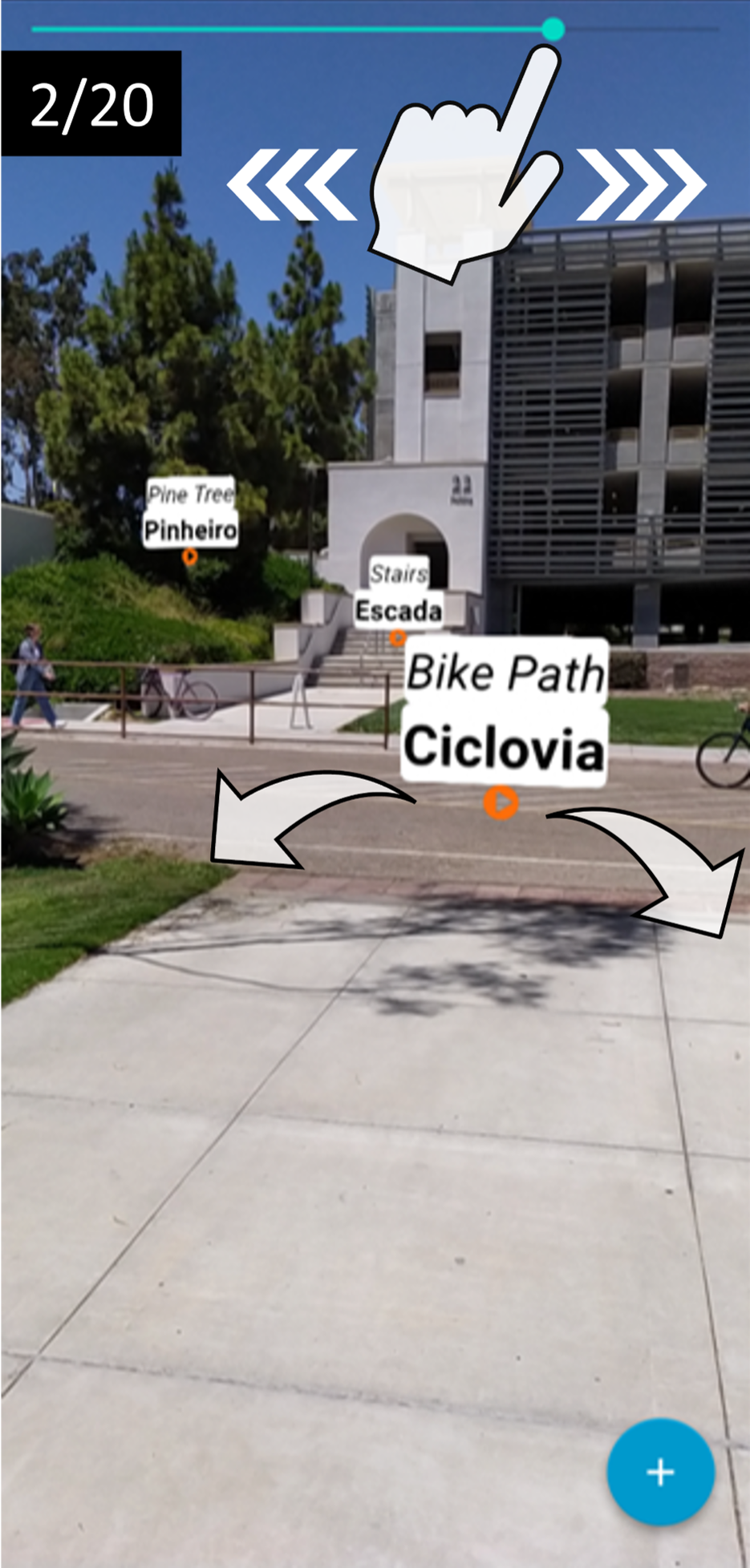}
         \Description[Adjusting tag positions]{ A sidewalk leading to a bike path as seen through the camera of a mobile device from the same perspective with the UI of the ARLang around the edges. On top of the bike path, there is an augmented reality tag with the word “Bike Path” and its Portuguese translation, “Ciclovia”. In the distance, there are two other tags on top of their physical counterparts, “Stairs” and “Pine Tree”. The tag counter is on the top left corner of the screen and says “2/20”. On the top edge, there is a slider UI element and an icon of a pointing finger, and around it, there are arrows pointing to the left and right, depicting the movement of the slider. Next to the “Bike Path” AR tag, there are two arrows pointing left and right curved around the camera depicting the sliders moving the AR tags around the Y-axis.}
         \caption{}
         \label{fig:slider_and_counter}
     \end{subfigure}
    \caption{
    (a) Left: Unplayed AR tag on top of a barbecue grill displaying ``Churrasqueira'', the noun in Portuguese, and ``Barbecue Grill'', the English translation. Right: The green border around the noun in Portuguese indicates the tag was played. Figure \ref{fig:ankiapp} shows the same content as a flashcard.
    (b) Left: Buttons on the screen's bottom-right corner load and unload the AR tags. Right: Buttons on the screen's bottom-right corner fold after loading AR tags. 
    (c) On the top edge of the screen, the slider is used to adjust AR tag rotation. The counter of played tags is shown below the slider. 
}
\end{figure*}

\section{ARLang System Design}
\label{sec:system_design}
We developed an outdoor mobile AR app called ARLang to investigate the objective of this research is to compare the effectiveness and user experience of outdoor AR labels with a non-AR method for learning foreign vocabulary. We aim to gain insights into the effectiveness of AR for language learning and use the findings to inform the design of future AR-based language learning tools. The prototype, ARLang is a native Android app powered by ARCore, Google's software development kit \footnote{\url{https://developers.google.com/ar}} which provides necessary features for grounding AR in the real world, such as mobile tracking and object anchoring. 
In 2021, Google reported having three billion Android users worldwide, indicating that mobile AR currently has a higher chance of being used by consumers than head-mounted AR \cite{verge2021android}.

The AR tags attached to physical objects, as seen in Figure \ref{fig:artag}, are the main UI elements of ARLang. A tag is composed of a stack of three elements. A circular orange play icon at the bottom of the stack indicates the tag can play audio content. The two elements on the top of the stack are white horizontal rectangular labels with rounded corners. The rectangular label shows the Portuguese noun in bold black font. The topmost label shows the English translation in italic black font. The font size is large enough to be visible on the farthest object. We manually defined the position of AR tags to match the real-world coordinates of their respective object in a way AR tags appear to users to be next to their real-world object correspondents. As the user moves around the outdoor environment looking for AR tags, they learn Portuguese nouns in the real-world context where those objects exist in a situated learning experience \cite{lave1991situated, brown1989situated, dunleavy2014augmented}. Users can tap on the tag to hear the pronunciation audio and look at the tag to read the word, check the spelling, and how it relates to the pronunciation, favoring a multimedia learning approach \cite{mayer2022multimedia}. After the user plays a tag, a green border appears around its Portuguese label to help users quickly identify tags they have already played. Users can play the same tag as many times as needed.

The auxiliary UI elements are arranged in screen coordinates instead of anchored to real-world objects as AR tags. Figure~\ref{fig:foldable} shows two foldable buttons on the bottom-right corner of the screen to load and unload the AR tags. The first button loads the tags into the current scene with respect to the camera position and orientation, and the second button removes all the tags from the scene. There is a slider and a counter at the top of the screen, as shown in Figure \ref{fig:slider_and_counter}. The counter on the top left corner of the screen provides feedback on how many tags the user has played. This component contributes to a game-like experience, motivating users to find all the words in the area. The slider on the top edge of the screen rotates all the AR tags about the y-axis of the global coordinate system to offset tracking instability and ensure tags are always in their correct positions. Tracking instability, or spatial drift, is a common issue in tracking-based AR caused by device movement or changes in lightning conditions that results in errors in the tracking algorithm and misplacement of AR components over time \cite{slocum2021drift, scargill2022drift}. The experimenter adjusted the slider before the start of the learning session and re-adjusted it when needed. Participants were instructed not to use this slider.

The final system design was the product of an iterative design process. We developed an initial prototype and conducted a pilot with volunteers to collect early feedback on the ARLang design. We collected important feedback that motivated changes in the following design iteration, such as increasing the font size to improve readability, adjusting AR tag placements for better correspondence with physical objects, changing from fading tags after tapping to highlighting them to avoid discouraging users playing a tag multiple times, adding a UI element to display the count of visited tags and the total number of tags for feedback on user progress in the study session, and editing audio for consistent volume and trimmed silence.

\begin{figure}[!ht]
\includegraphics[height=6.5cm]{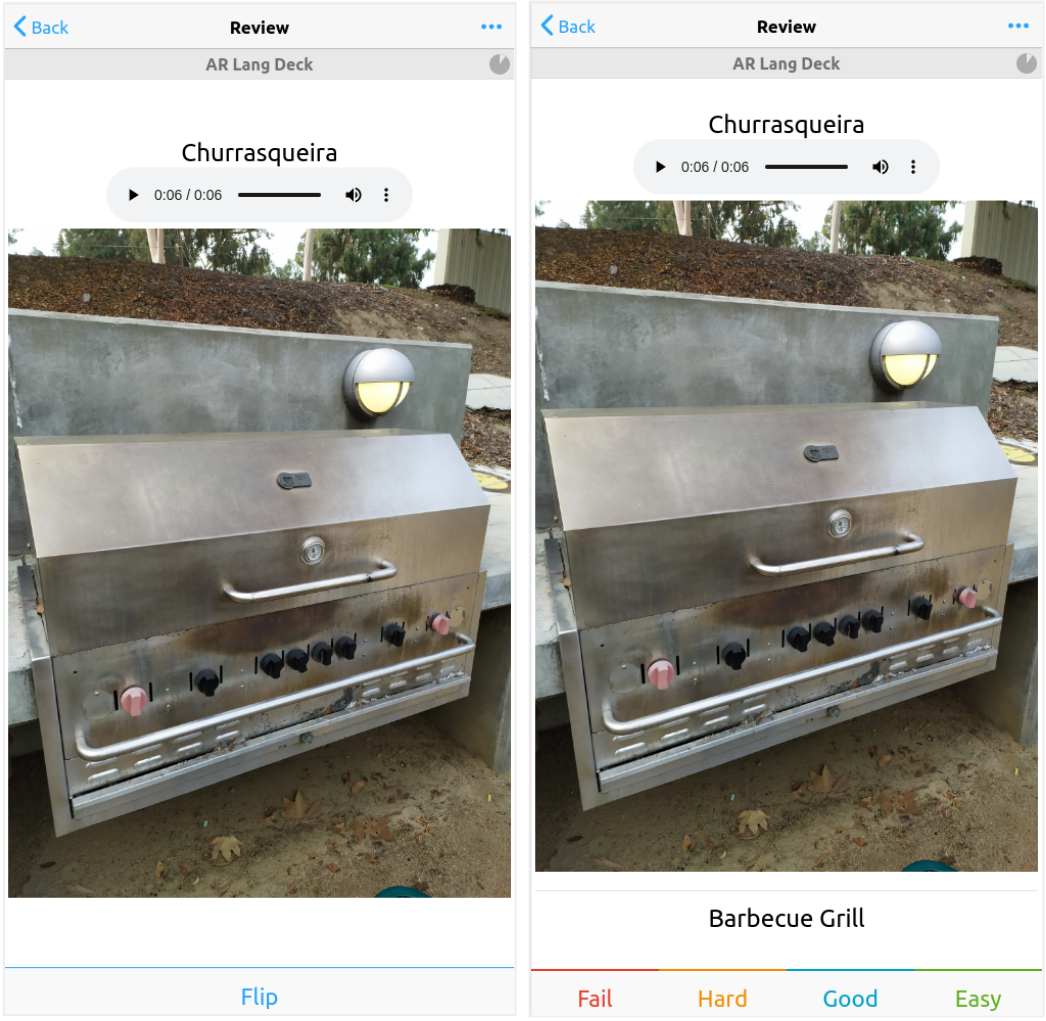}
\Description[AnkiApp Flashcards]{On the left, there is a screenshot of an AnkiApp flashcard. From top to bottom, there is a label that says “Churrasqueira”, an audio player, a photo of a barbecue grill, and a button “Flip”. On the right, there is the same image except for a label that says “Barbecue Grill” underneath the barbecue grill photo and the buttons “Fail”, “Hard”, “Good”, and “Easy” instead of the “Flip” button.}
\caption{Left: Front face of a flashcard with Portuguese written noun, Portuguese pronunciation audio, and image of a barbecue grill. Right: Back face of the same flashcard with Portuguese written noun, Portuguese pronunciation audio, an image of a barbecue grill, English translation, and difficulty scale. Figure \ref{fig:artag} shows the same content as an AR tag.}
\label{fig:ankiapp}
\end{figure}

\section{User Study}
\label{sec:user_study}

We conducted a between-subjects user study with 44 participants divided into two groups. Each group performed the same Portuguese vocabulary learning task under two distinct conditions. While more expensive and time-consuming, we chose a between-subject experiment design to minimize carryover and fatigue effects between conditions. A within-subject design could have introduced confounding variables such as learning patterns in Portuguese translations and physical fatigue from the outdoor condition. The test group used ARlang and the control group used AnkiApp, a free flashcard mobile application \cite{ankiapp2022flashcards}. As seen in Figure \ref{fig:ankiapp}, the front face of a flashcard displays the Portuguese noun and the pronunciation audio. Once the users memorize the front, users can flip the flashcard to see the written English translation. We chose flashcards for the control group based on prior work in language learning \cite{ibrahim2018arbis}. This user study aims to address the following research objectives:
\begin{itemize}
\item Compare the effectiveness in terms of L2 learning outcomes of an outdoor AR tool versus a non-AR method.
\item Compare the user experience provided by an outdoor AR tool versus a non-AR method for L2 learning.
\end{itemize}

\subsection{Participants}
The study was approved by our office of research (protocol number: 9-22-0513) and participants were compensated \$20 for their time. Before starting the experiment, all participants signed an informed consent form and filled out a pre-study survey questionnaire. Of the 44 participants, one participant identified as non-binary, 18 as female, and 25 as male. The mean age was $21.98$ years with a standard deviation of $4.05$. Regarding familiarity with AR and flashcard apps, 26 participants affirmed having ``None'' or ``Very little'' experience using AR applications, and 20 declared having ``A lot'' or ``Extensive'' experience with mobile flashcard applications. Participants answered the questions using a five-point Likert Scale (None=1, Extensive=5). None of the participants declared having any familiarity with the Portuguese language. Seven participants stated familiarity with other Romance languages, namely Spanish and French.

\begin{figure*}[!ht]
\includegraphics[width=0.8\textwidth, height=5.5cm]{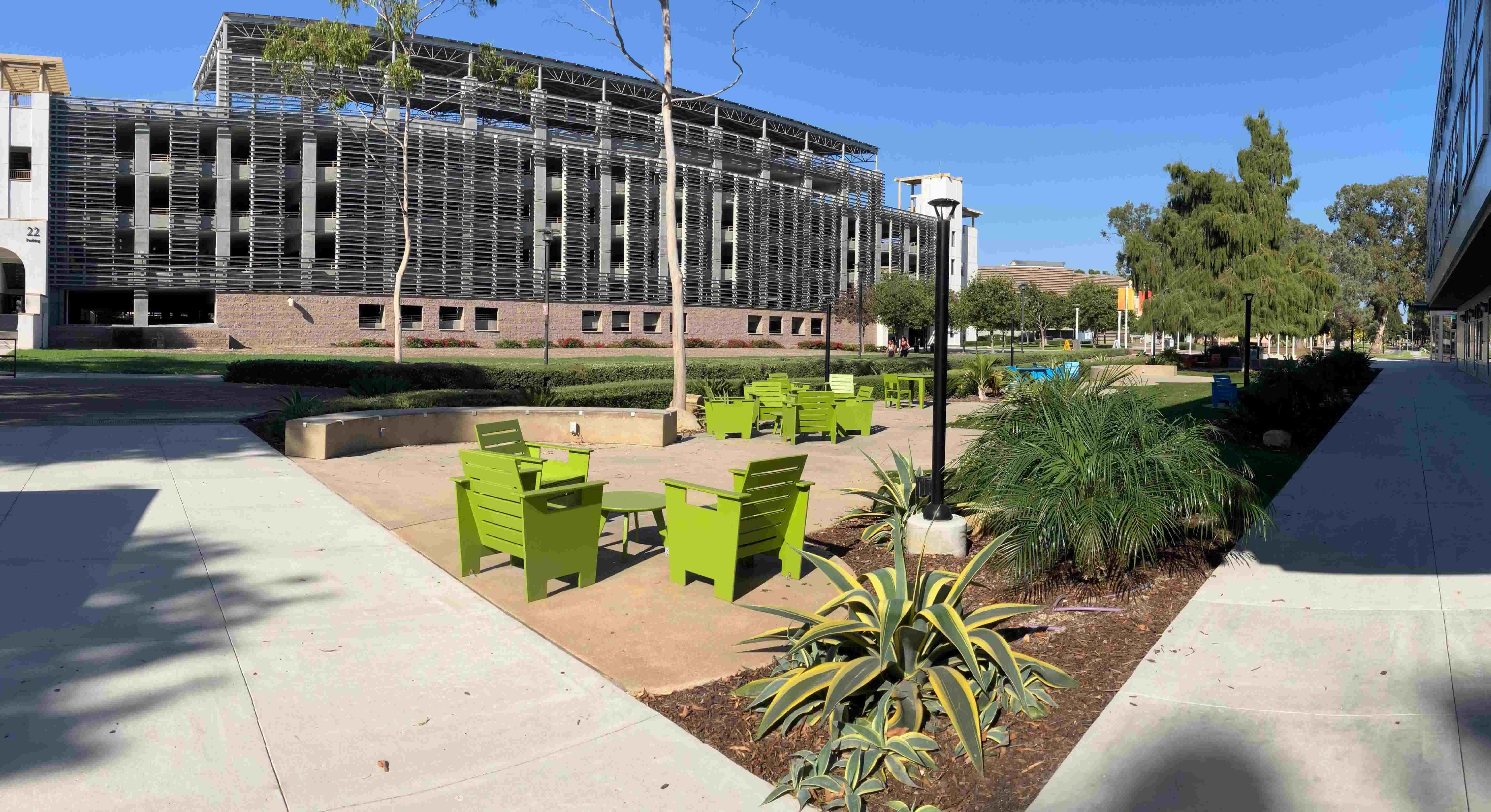}
\Description[Study Area Panorama]{A panorama image of a spacious rectangular area with several groups of green chairs and small tables scattered around. Bushes are situated next to the area, while a wide sidewalk surrounds it. Further in the distance, one can spot a bike path and a multi-level parking structure.}
\caption{
View of the outdoor ARLang study area showing pedestrian sidewalks, seating spaces, and campus buildings.
}
\label{fig:panorama}
\end{figure*}

\subsection{Experiment Setup}

We randomly split participants into two groups of 22 each. The test group used ARLang, and the control group used AnkiApp, a free flashcard mobile application. The test group performed the learning task in an outdoor area (Figure \ref{fig:panorama}) located on a walking route between our campus and a local student neighborhood between $9$ AM to $5$ PM in daylight. All outdoor participants encountered the same weather conditions. The test group study location had the expected outdoor elements, such as the presence of other people and general outdoor noise in a pedestrian area. All users in the ARLang group used the same mobile device during the learning sessions: an ARCore-compatible smartphone (6.3 inches display, 1080 x 2340 pixels, Qualcomm SDM660 Snapdragon 660, Adreno 512, Android 10). The flashcard group (control group) performed the learning task in a university classroom between $9$ AM to $5$ PM under artificial light conditions. Participants in this group used their personal mobile devices. We consider all participants in this study to have similar levels of familiarity with mobile devices and mobile apps. 

We chose Portuguese because there is limited previous work on AR tools for learning Portuguese as a second language. In addition, there is a small Portuguese speaker community on our campus, which helped lower the chances of finding  participants with Portuguese familiarity. Lastly, one of the co-authors is a native Portuguese speaker who helped facilitate design choices and analysis. We consulted a teacher of Portuguese as a second language to understand how many words and how long the learning session should take. Based on their suggestions, we set the study task to learn $20$ words in $10$ minutes. These parameters were subsequently validated in a pilot study with $5$ representative users. 

We selected $20$ static objects visible while walking in the chosen outdoor area to derive the nouns used in the study. Figure \ref{fig:als} shows the nouns in English (first row) and their respective Portuguese translation (second row). For each word, both participant groups had access to the following content: text in Portuguese, text in English, audio of the whole Portuguese native pronunciation, and audio of the syllable-by-syllable Portuguese native pronunciation. In addition to the multimedia content, the test group had physical access to the actual object referred to by the Portuguese word, and the control group had access to a picture of the same object. Both experimental conditions followed the same time constraint of $10$ min. It is reasonable to consider different time limits for each experimental group because the ARLang users will need to walk between AR tags and flashcards. Users can tap on the screen to move to the next word. However, keeping the same time constraint allows us to directly compare results across groups and isolate the effects of situated learning \cite{brown1989situated, lave1991situated, dawley2014situated} and associating information to physical space \cite{fujimoto2012memorization, rosello2016nevermind}, that are known to favor learning and memorization as stated in previous literature. As the results show, $10$ min was enough time for all participants in both conditions to study every word at least once, discarding the possibility that a user encountered a word in the quiz that they never had the chance to learn. 

\begin{figure*}[!ht]
\includegraphics[width=0.8\textwidth, height=5.5cm]{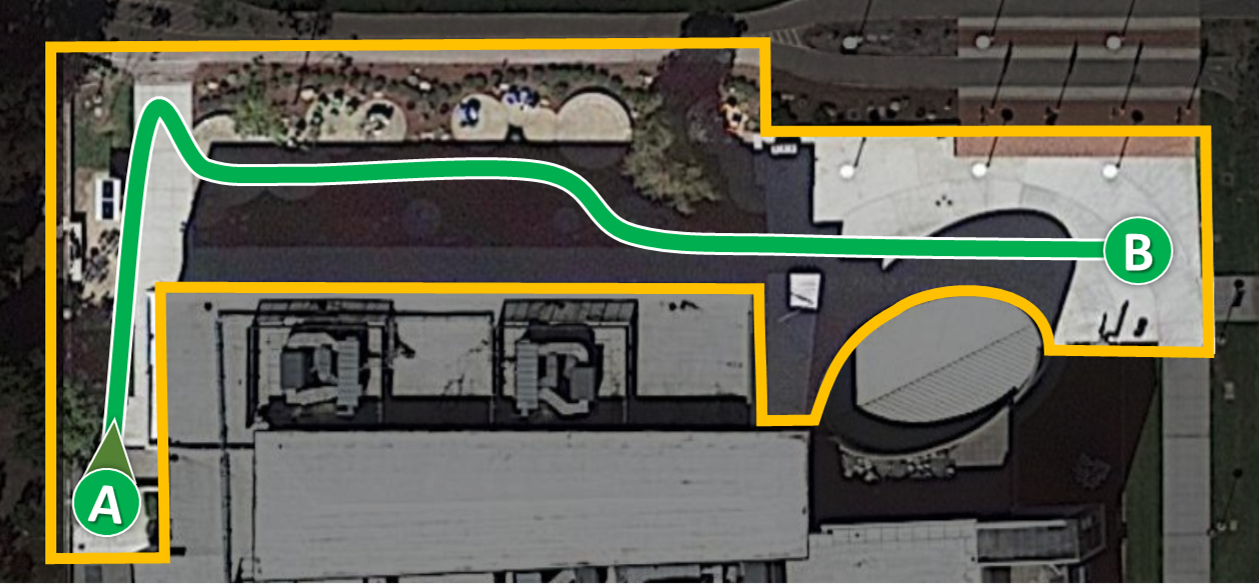}
\Description[Study Area Trajectory]{A bird’s-eye view of the rectangular with scattered chairs and tables surrounded by vegetation. On the bottom, there is a rooftop of a rectangular. On the top, there is a bike path. The walkable region of this area is highlighted and there are two points, A and B, marking the beginning and the end of a path. The points are connected by a thick line depicting a trajectory that goes from the left extreme to the right extreme of the area.}
\caption{
Bird's-eye view of the location where the ARLang group conducted the user study. The orange border bounds the approximate study area and the green line marks the walking path along which AR objects are labeled, starting at point A. 
}
\label{fig:srb}
\end{figure*}

\subsection{Experiment Procedure}

The ARLang group (test group) performed the learning task in the outdoor study area. All participants used the same Android mobile device and headphones provided by the experimenter. Participants started the study session at the same spot (Figure~\ref{fig:srb}, point A) and in the same pose, i.e., they all faced the same way holding the phone at eye height. The experimenter asked participants to stay within the $900m^2$ study area at all times by pointing at landmarks and asking them not to go beyond those. Within this area, we defined a $135m$ long path along which we placed the AR object labels. Notifying participants of the area bounds prevented them from straying too far from the AR object labels and helped them stay safe on a flat walking surface as they looked at the world through the phone's camera.
Participants were free to visit tagged objects in any order by walking around as they chose. Figure \ref{fig:srb} shows a bird's eye view of the outdoor location with the study area bounds, the initial participant position and orientation, and the walking path. The experimenter demonstrated how to use the app and only started the study when participants expressed being comfortable with the interface and ready to start. The experimenter tapped the ``start'' button to load the tags and set a $10$ min timer. To ensure safety, the experimenter followed the participants, staying approximately $3$ meters behind them. Questions about the precise location of the AR tags were not answered. When a trial reached the $10$-minute mark, participants returned the mobile device to the experimenter, who tapped on the ``end'' button to end the trial and remove the AR tags from the scene.

\subsubsection{Distractor Task} 
After the learning task and before assessing recall, participants from both groups watched a video collection of comic dog clips knowing they would have to answer a question about them. The publicly available YouTube video was $2$ minutes and $21$ seconds long \footnote{\url{https://youtube.com/watch?v=kMhw5MFYU0s}}. After watching the video, they answered the question: ``Which dog clip was your favorite from the previous video?'''. The objective of this seemly unrelated task was to make participants use their working memory on a distractor task to prevent rehearsal of the words before the recall task \cite{peterson1959short}. This strategy is considered equivalent to a delayed recall test. Prior learning studies have successfully used this design to avoid losing participants due to the time gap \cite{hirst1988distractor}.

In the last part of the experiment, users answered a quiz with two types of questions, En-Pt translations and Pt-En translations. We counterbalanced the quiz by randomly assigning participants to one of two versions to help mitigate directional bias \cite{gaito1961repeated}. The first version of the quiz starts with 10 En-Pt questions followed by 10 Pt-En questions. The second version of the quiz switches the order of the two question sets. Each vocabulary word was used only once in either of the two question sets.
Participants answered the questions by typing in a free text field. The experimenter oriented participants to answer as precisely as they could recall or input ``I don't know'' if they did not remember anything. After the quiz, participants completed a post-study questionnaire regarding their subjective experience of the learning task and spatial abilities.

\subsection{Evaluation Metrics}

We computed the overall short-term recall score as the ratio of total correct answers from the two question sets in the quiz. We also calculated the short-term recall scores considering only the Portuguese-to-English (Pt-En) question set and the English-to-Portuguese (En-Pt) set. Participants answered the digital quiz in free-text format. Answers that perfectly matched the correct translation were worth 1 point. To assign partial credit to answers with typos or minor spelling mistakes, we adopted Adjusted Levenshtein Score (ALS) (Equation \ref{eq:als}) \cite{levenshtein1966binary, ibrahim2018arbis}. An ALS of $0$ means the provided answer is as far as possible from the correct answer. An ALS of $1$ means the provided answer perfectly matches a correct answer. This metric allowed us to assign partial credit for answers missing diacritical marks, very common in Portuguese, e.g., ``túnel''(Pt), ``lâmpada''(Pt), and ``ciclovía''(Pt).

More than memorizing pairs of Portuguese and English words, we want to measure the participant's understanding of the signified concept. For this reason, we compare the provided answers with a list of accepted synonyms. The accepted synonym list was created by compiling participants' answers and consulting Portuguese and English native speakers (2 of each) to indicate which of the provided answers were synonyms of the original correct answer. The final score of a question is the maximum ALS between the input answer and every accepted synonym of the correct answer. As an example, consider a participant misspelled the answer ``barbecue grill'' as ``barbec grill'' when asked for the English translation of ``churrasqueira''(Pt). The accepted synonyms for the translation of ``churrasqueira''(Pt) are ``barbecue grill'' and ``barbeque grill''. If the ALS scores are $ALS(\text{``barbec grill''}, \text{``barbecue grill''}) = 0.86$ and $ALS(\text{``barbec grill''}, \text{``barbeque grill''}) = 0.78$ then the score for that particular answer would be $0.86$.

\begin{table}[h]
    \begin{equation} 
        \label{eq:als}
        \textrm{Score}(w,\hat w) = 1 - \frac{\textrm{min}(Lev(w,\hat w), |\hat w|)}{|\hat w|}
        \end{equation}
    \vspace{0.2cm}
    \begin{center}
        \begin{tabular}{ll}
            $w$: & \textrm{attempted answer} \\
            $\hat w$: &\textrm{correct answer} \\
            $Lev$: &\textrm{Levenshtein distance} \\
            $|\hat w|$: &\textrm{length of the correct answer $\hat w$}
        \end{tabular}
    \end{center}
\end{table}

We measure learner experience by asking participants custom questions about their motivation to learn, the difficulty level of the words, the effort the participants put in, how much they enjoyed the experiment, and whether they would like to learn a new language using the same tool in the future. Participants answered these questions on a 5-point Likert scale ($1$=Strongly Disagree and $5$=Strongly Agree). Additionally, we used three questions from the Santa Barbara Sense of Direction Scale \cite{hegarty2002development} to assess participants' environmental spatial abilities, as this metric could help us to identify disparities in the performance of the AR group due to movement in space outdoors. The specific questions were ``I have a poor memory for where I leave things'', ``I can usually remember a new route after I have traveled it only once'', and ``I don’t have a very good `mental map' of my environment''. Participants answered these questions on a 7-point Likert scale (1=Strongly Disagree and 7=Strongly Agree).

\section{Results}
\label{sec:results}
This section presents the statistical methods used to analyze the study data. We start by characterizing the participant groups to conclude they are statistically similar. The study results do not indicate a significant difference in short-term recall scores between the two groups. However, regarding learner experience and preference, we see a significant difference in favor of the ARLang group.

\subsection{Characteristics Across Groups}
We conducted independent sample two-tailed t-tests comparing the following features between the two study groups based on data from the pre-study questionnaire: class level, grade point average (GPA), age, gender, number of languages understood proficiently, interest in other languages, proficiency in Portuguese, familiarity with AR technology, and familiarity with flashcard applications. The t-test results show both groups to be equivalent across all features 
(\textit{P}$<.15$). We used a Chi-square test to determine whether gender was equivalent across groups, which was also a non-significant difference between them, $\chi^2$ (df=$2$, N=$44$), \textit{P}$=.084$. We compared the number of participants proficient in a Romance language (French or Spanish) since the learning task could have been easier for participants familiar with one of those languages. A Chi-square test indicated both groups were equivalent in this aspect too, $\chi^2$ (df=$1$, N = $44$), \textit{P}$=.41$. These findings support the equivalence of the two groups of interest in all the considered factors.

\begin{figure*}[!ht]
\includegraphics[width=0.9\textwidth]{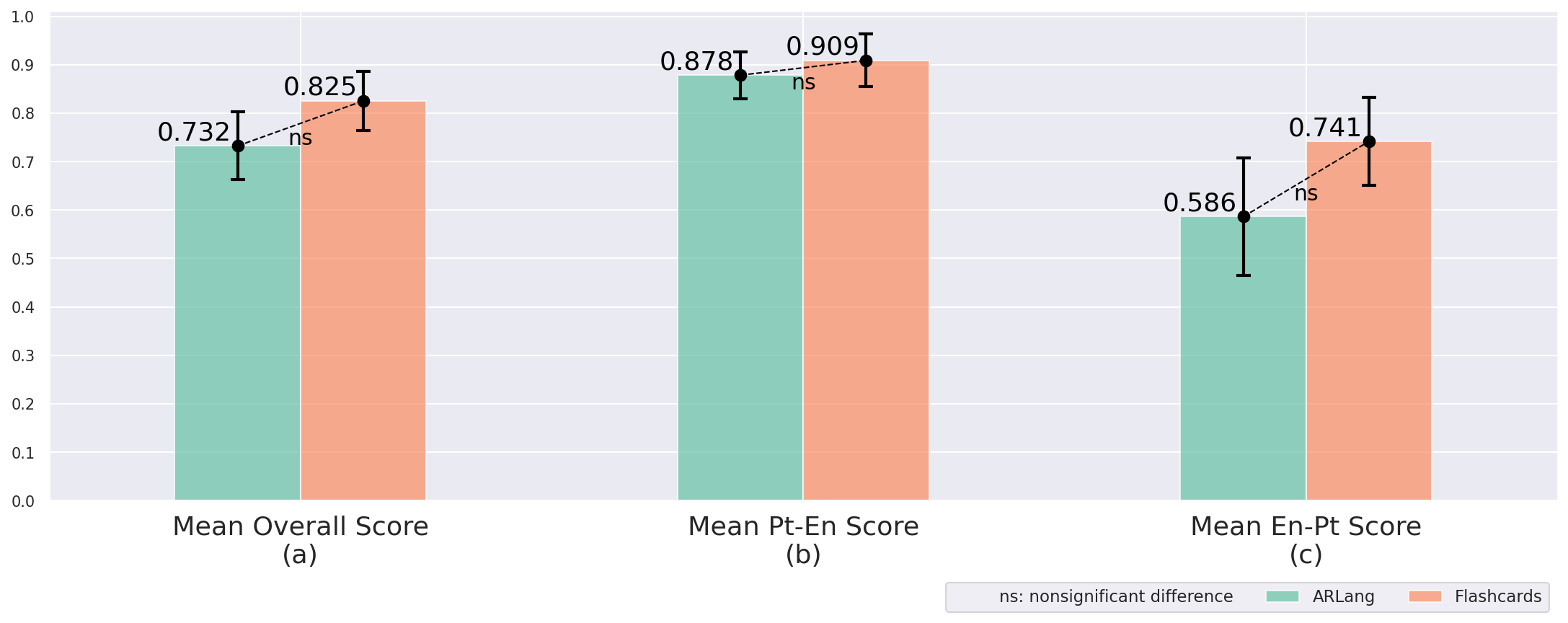}
\Description[Bar plot of Mean Scores]{Three bar plots, one for each section-wise breakdown of the quiz, with two bars each, one bar for each condition. In all the bar plots, the shorter bar is ARLang. There are dashed lines connecting the peaks of each bar of a pair showing the difference in the scores and a label ``ns'' meaning that there is no statistically significant difference between the values. There are error bars around the peak of each bar.}
\caption{(a) Mean Overall Scores of each experimental group with their $95\%$ confidence interval. (b) Mean Pt-En Scores of each experimental group with their $95\%$ confidence interval. (c) Mean En-Pt Scores of each experimental group with their $95\%$ confidence interval.}
\label{fig:scores}
\end{figure*}

\begin{table}[!ht]
\Description[Score Statistics]{The table displays the scores and statistics of ARLang and Flashcards conditions. The rows show the section-wise breakdown of the quiz, including "Overall", "Portuguese-to-English", and "English-to-Portuguese." The score values are presented as floating point numbers, with three decimals, ranging between 0 and 1. A 95\% confidence interval is provided next to the scores. The final column showcases the U-statistics and P-values of the Wilcoxon rank-sum.}
\caption{P-values and U statistic of a two-sided Wilcoxon rank sum test comparing score distributions of ARLang and Flashcards groups.}
\begin{tabular}{@{}rccrl@{}}
\toprule
 & ARLang & Flashcards & \multicolumn{2}{c}{Wilcoxon rank sum} \\ \midrule
Overall & $0.732\pm0.070$ & $0.825\pm0.061$ & \textit{U}$=1.79$ & \textit{P}$=.07$ \\
Pt-En & $0.878\pm0.049$ & $0.909\pm0.054$ & \textit{U}$=1.16$ & \textit{P}$=.24$ \\
En-Pt & $0.586\pm0.121$ & $0.741\pm0.091$ & \textit{U}$=1.88$ & \textit{P}$=.059$ \\ \bottomrule
\end{tabular}
\label{tab:scores}
\end{table}

\subsection{Overall Scores}
Considering both En-Pt, and Pt-En, the flashcard group (control) achieved a mean score of $0.825$ within a $95\%$ confidence interval $[0.764, 0.886]$. The ARLang group (test) achieved a mean score of $0.732$ within a $95.0\%$ confidence interval $[0.662, 0.802]$. Figure \ref{fig:scores}a shows the mean overall scores for each group with their respective $95\%$ confidence intervals. Shapiro-Wilk test rejected the normality of the score distribution for the flashcard group (\textit{P}$=.02$) and did not reject the normality of the score distribution for the ARLang group (\textit{P}$=.87$). Since the score distribution in the flashcard group is not normal, we compared the score distributions of each group using a two-sided Wilcoxon sum rank test. The difference between the group's overall scores is non-significant (\textit{P}$=.07$) (Table \ref{tab:scores}).

\subsection{Portuguese-to-English (Pt-En)}
Considering only the Pt-En questions, the flashcard group achieved a mean score of $0.909$ within a $95\%$ confidence interval $[0.854, 0.963]$. The ARLang group achieved a mean score of $0.878$ within a $95\%$ confidence interval $[0.829, 0.927]$. Figure \ref{fig:scores}b shows the mean Pt-En scores for each group with their respective $95\%$ confidence intervals. Shapiro-Wilk tests rejected the normality of the Pt-En score distribution for the flashcard group (\textit{P}$<.001$) and also rejected the normality of the Pt-En score distribution for the ARLang group (\textit{P}$=.02$). Since the score distribution from both groups is not normal, we compared the score distributions of each of them using a two-sided Wilcoxon sum rank test. The difference between the group's scores on the Pt-En problem set is non-significant (\textit{P}$=.24$). See Table \ref{tab:scores} for a summary. The first row of Figure \ref{fig:als} shows the boxplots of the Portuguese-to-English ALS scores of each noun.

\subsection{English-to-Portuguese (En-Pt)}
\begin{figure*}[!t]
\includegraphics[width=\textwidth]{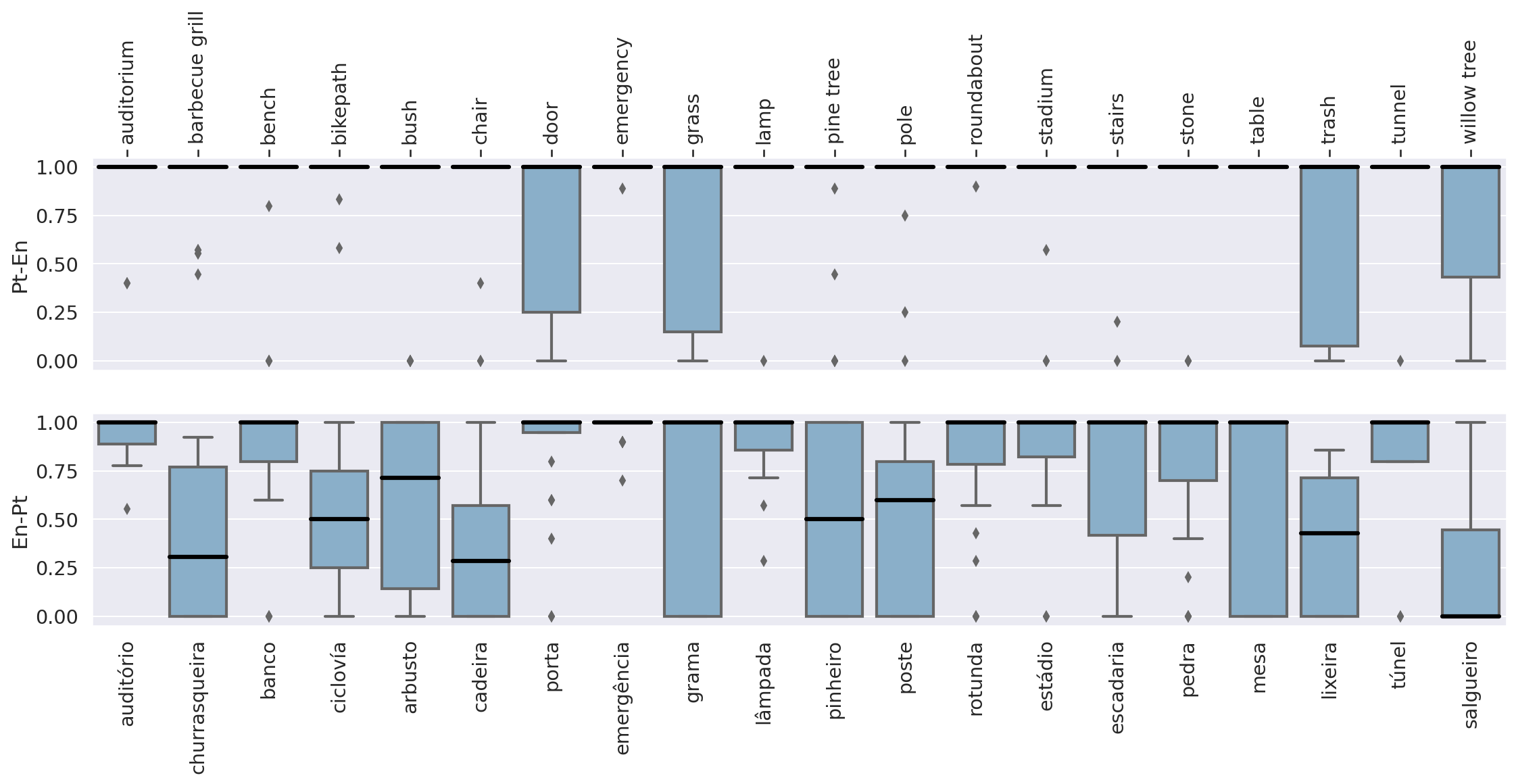}
\Description[Boxplot of scores]{Two bar plots of the score distribution. There are 20 boxes on each boxplot, one per word. The boxplot on the top shows the score distributions of the Portuguese-to-English section of the quiz and the boxplot on the button shows the English-to-Portuguese score distribution. Portuguese-to-English score distributions are more skewed to 1 than the English-to-Portuguese.}
\caption{Boxplot of the ALS scores from all participants in each of the 20 nouns in the study. The Portuguese-to-English (Pt-En) scores are in the first row, and the English-to-Portuguese (En-Pt) scores are in the second row.}
\label{fig:als}
\end{figure*}

Considering only the En-Pt questions in the quiz, the flashcard group achieved a mean score of $0.741$ within a $95.0\%$ confidence interval $[0.650, 0.832]$. The ARLang group achieved a mean score of $0.586$ within a $95\%$ confidence interval $[0.464, 0.707]$. Figure \ref{fig:scores}c shows the mean En-Pt scores for each group with their respective $95\%$ confidence intervals. Shapiro-Wilk tests rejected the normality of the En-Pt score distribution for the flashcard group (\textit{P}$=.02$) but did not reject the normality of the ARLang group (\textit{P}=$.46$). Since the score distribution from the flashcard group is not normal, we compared the score distributions of each group using a two-sided Wilcoxon sum rank test. The difference between the group's scores on the En-Pt problem set is non-significant (\textit{P}$=.059$). See Table \ref{tab:scores} for a summary. The second row of Figure \ref{fig:als} shows the boxplots of the English-to-Portuguese ALS scores of each noun.

\subsection{Experience with the Lessons}
We found that motivation, difficulty, effort, and enjoyment were similar across groups (\textit{P}$>.11$). The question ``I would like to learn a new language this way in the future'' showed a significant difference between the groups, t($42$)$=2.04$, \textit{P}$=.048$, with those in the ARLang group (M$=4.23$, SD$=.92$) agreeing with the statement more than those in the flashcard group (M$=3.64$, SD$=1.00$), which indicates a higher preference for ARLang over the flashcard alternative. Additionally, there were no significant differences between the groups in any of the Santa Barbara Sense of Direction Scale questions (\textit{P}$>.21$). We share below the open-ended feedback participants provided to help contextualize our analysis.

\subsubsection{Positive Sentiments} 
Participants from both groups were positive about the study session, but more so in the ARLang group. \textit{``It was fun to learn the words with both image and pronunciation available.''} (P51, Flashcards). \textit{``It was a lot of fun! (...)''} (P03, ARLang). Some participants of the ARLang group associated the positive experience with the possibility of interacting with the outdoor context. \textit{``Cool way of interacting with the environment''} (P16, ARLang). \textit{``Fun to find words (...)'' (P05, ARLang)}. Using tools that learners prefer can help improve motivation which has been shown in several studies to be associated with better learning outcomes \cite{liu2012measuring,steinmayr2009importance}.

\subsubsection{Challenges of Outdoor AR} 
\label{subsub:challenges}
Participants of the ARLang group reported minor issues with registration drift and occlusion, known common problems in AR.
\textit{``(words) don’t always stick with the object they mean''} (P05, ARLang). \textit{``The AR got out of focus during the last part of the study but it was very good otherwise''} (P15, ARLang). \textit{``Experienced some registration issues, although it corrected itself. Occlusion was not always perfect.''} (P17, ARLang). One of the participants highlighted the impact of pedestrians present in the outdoor location during the experiment. \textit{``I felt like people thought I was recording them which felt weird. An indoor setting may have been better''} (P24, ARLang). This comment, in particular, exemplifies how complex social aspects of technology use in public can play a role in its acceptance and use for learning in the outdoor context \cite{wolf2014lifelogging}.

\subsubsection{Recognition and Production Tasks} 
Participants reported recognizing words in Portuguese to be easier than recalling them for writing based on their English translations. Recognizing words encountered previously is easier than actively writing based on a cue \cite{neely1977semantic}. \textit{``I can recognize them pretty well but I need more time to learn the spelling.''} (P43, Flashcard). \textit{``Easy to provide Portuguese to English but not English to Portuguese''} (P08, ARLang). \textit{``(...) I do not remember the spelling.''} (P37, Flashcards). Overall, this is understandable since in general, recognition is easier than recall \cite{haist1992relationship} 

\subsubsection{Improving ARLang} 
Participants provided some suggestions to help improve ARLang. Some suggested combining ARLang with real-time object recognition to display AR labels in any location.
\textit{``(...) It would be great if it can be paired with object detection and the app can decide which word to show (...)''} (P04, ARLang).

One participant mentioned combining both tools in a mixed learning practice where learners could alternate between mobile flashcards and the AR learning experience. \textit{``I think this would be most helpful if I spend time reviewing the vocabulary beforehand to at least be familiar with the words, then try to learn them in AR.''} (P17, ARLang). \textit{``maybe have a test mode in the app or a way to see the items you visit with a picture to revisit them.''} (P01, ARLang).

\section{Discussion}
\label{sec:discussion}

We initially expected that contextualizing L2 learning experiences outdoors with AR would positively impact learning outcomes based on learning theories \cite{lave1991situated, brown1989situated, mayer2022multimedia}, and empirical evidence in related work \cite{liu2013essays, hautasaari2019vocabura, ibrahim2018arbis,
weerasinghe2022vocabulary, rosello2016nevermind}. However, our results show that while users preferred vocabulary learning in outdoor AR over flashcards, they did not indicate improvement in short-term recall. Here we discuss our results in light of prior work.

\subsection{Short-term Recall}

\subsubsection{Environmental Factors} 
Exploring language learning in a real-world context 
is a central part of this work. Since the outdoors is an uncontrolled environment, learners in the ARLang group likely experienced several potentially distracting factors, including splitting attention between the surroundings and the screen to navigate around obstacles and people as well as managing noise. While headphone use while walking is common on a college campus, we suspect the impact of the noise distraction may have been a bit lower than the visual distractions, the novelty of AR, and the social implications of constantly holding a phone up to view the world. Prior work agrees on some of these distracting factors hindering learning \cite{lave1991situated, brown1989situated, mayer2022multimedia}. These distractors have been largely missing in prior AR language learning studies performed indoors that have shown improved recall using AR compared with flashcards \cite{ibrahim2018arbis, weerasinghe2022vocabulary}. Another factor that may have negatively impacted the learning performance outdoors was the higher physical effort by the ARLang group. A recent study analyzed electroencephalography (EEG) P3 measurements demonstrating that performing a task while walking induces higher cognitive load than when sitting down \cite{sweller1994cognitive, swerdloff2022eeg}. Social factors associated with public outdoor contexts could have played a role in impeding learning outdoors. As mentioned in Section~\ref{subsub:challenges}, social discomfort could have distracted users from the learning task and reduced their performance. 

\subsubsection{Subjective Factors} 
The lack of familiarity with the mobile AR interaction paradigm might have been another factor that impacted the learning performance. In the ARLang group, 63\% of the participants declared having ``None'' or ``Very little'' experience with AR. In the flashcard group, only $23$\% of the participants reported having ``None'' or ``Very little'' experience with flashcard apps. Even though the experimental groups were statistically equivalent regarding familiarity with AR and flashcard apps, the different levels of familiarity might have contributed to the results. 

\subsubsection{Technological Factors}
\label{sub:tech_factors}

Our results contrast with previous work on mobile AR L2 learning that observed a higher learning performance in the outdoor condition \cite{liu2013essays, hautasaari2019vocabura}. We explain these divergences pointing to differences in the evaluation method employed by Liu et al. \cite{liu2013essays} that did not measure short-term recall and differences in the interface, as their application uses AR elements as shortcuts for a 2D interface that displays the learning content instead of directly registering it in space. Hautasaari et al. \cite{hautasaari2019vocabura} suggested combining audio and visual elements for outdoor L2 learning experiences as we did in ARLang. Our results show that this combination might not be enough to positively impact learning performance. The design of an effective AR experience for L2 learning outdoors needs further exploration. Subsection~\ref{subsub:challenges} mentions users reporting some registration issues. Although our interface includes a slider to correct drift as described in Section~\ref{sec:system_design}, drift may have distracted users from the learning task. Another possible factor that might have influenced the results is that ARLang is a proof-of-concept app. Although it has all the necessary features for the study task, it is not as polished as the flashcard application. 

\subsection{Higher Preferability}

As seen in the results, participants who used ARLang wanted more lessons in AR than those who tested the flashcard app. This preference indicates the potential for a deeper and longer engagement with the study material, which is highly desirable for language learning. These findings corroborate observations from previous research \cite{cai2017magnetic, santos2016augmented}. The combination of higher preference and low familiarity with AR can suggest a novelty effect \cite{bursali2019effect}. We attempted to mitigate this by giving participants as much time as needed to get familiar and comfortable with the interface.

\subsubsection{Indoors vs. Outdoors}
As our results demonstrate, designers of AR learning experiences need to consider differences in indoor and outdoor spatial contexts in the design process. Although learning opportunities occur in both, they have distinct environmental conditions that can impact the learning experience. Indoor spaces (e.g., classrooms, auditoriums, laboratories) are often more stable and controllable. In contrast, outdoor spaces (e.g., parks, streets, campuses) are less predictable due to the presence of visual clutter, high noise levels despite headphone use (e.g., traffic, other people talking, skateboards), changes in weather and lighting conditions (e.g., cloud cover alternating with bright sunlight), obstacles (e.g., uneven sidewalk, overgrown shrubs), and the presence of multiple dynamic entities (e.g., pedestrians, cyclists, cars). These differences can affect user behavior and consequently their learning experience. For example, walking outdoors with a mobile device decreases visual acuity, and experts recommend increasing font size by 20\% to compensate for this impact \cite{conradi2014analysis}.

Overall, our study shows that mobile AR language learning was preferred over flashcards. As prior work shows, understanding learner preferences can help support instructional design that enhances engagement, which is a critical motivator \cite{lucardie2014impact}, and improves receptivity to complex topics \cite{ilin2022role}. Our AR interface was significantly preferred over the flashcard alternative. This finding indicates potential for exploring the design of newer AR interfaces that integrate virtual content grounded and contextualized with real-world objects to optimize usability for tackling the unique challenges of learning in the outdoor environment. 

\section{Design Considerations}

The design and evaluation of ARLang yielded design considerations that we believe would be beneficial to the community interested in developing outdoor AR learning experiences.

\paragraph{Consider Contextual Content Generation}
Our current proof-of-concept focuses on vocabulary tags manually placed in a specific location. To reduce manual setup and add more depth to learning material, we recommend leveraging image-to-text AI models\cite{tewel2022zerocap}. Additionally, it would be possible to create prompts for large-language models~\cite{chiang2023vicuna} based on descriptions of the environment to suggest conversation topics based on the immediate surroundings and explore verbal skills with social interaction in outdoor spaces, e.g., standing in front of a restaurant or walking on the beach.

\paragraph{Provide Multimodal Content}
Visual AR components may be more suitable for engaging stationary learners, such as users sitting in a park or eating a meal outdoors. On the other hand, spatialized audio-based AR interactions when moving could help minimize splitting attention between navigation and interacting with the mobile AR interface. This dual approach could even help to reduce the social discomfort of walking among other people while holding the camera up, as reported by (P24, ARLang). The combination of an audio and visual interface can help to avoid crowding the screen space and enable users to focus on their primary tasks.

\paragraph{Consider Tracking Instability}
Tracking instability can result in virtual objects drifting and negatively impacting the user experience. However, combining GPS readings with visual-inertial data can improve the accuracy of label placement, which is now possible with modern AR frameworks \footnote{https://developers.google.com/ar/develop/c/geospatial/geospatial-anchors}. It is advisable to avoid relying solely on high registration precision and instead consider adapting the size of visual elements to maintain a visual relationship with physical objects despite some degree of drift.

\section{Limitations}

Short-term recall measurement following a distractor task is directly associated with learning effectiveness, but it does not describe the phenomena entirely. Measuring delayed recall (e.g., between 4 and 7 days) would also be valuable to evaluate the AR experience's effects on knowledge retention. Follow-up work should consider measuring delayed recall as some previous results demonstrate better knowledge retention in learners who used AR \cite{ibrahim2018arbis, hautasaari2019vocabura, rosello2016nevermind}. To better understand the role of distractions present in the outdoor context in impeding learning with an AR tool, future work could consider collecting cognitive load measures through physiological sensing and subjective measures. Prior research has evaluated L2 learning outcomes using mobile~\cite{santos2016augmented, draxler2020augmented} and HMD~\cite{ibrahim2018arbis, weerasinghe2022vocabulary} AR tools. However, we also consider that directly comparing an AR indoor condition with an AR outdoor condition can help determine the effects of outdoor distractions on an L2 learning task.

We acknowledge the limitations induced by the conditions needed to create a controlled environment that differs considerably from real-world scenarios and could have potentially introduced bias in the data collected. To address this limitation, we plan to conduct an in-the-wild experiment where both conditions will be deployed in real-life situations, and quantitative and qualitative data will be collected using interviews and surveys. We encourage readers to consider the results carefully as the sample size of 44 participants in the between-subject evaluation may not be sufficient to demonstrate statistical significance.

\section{Conclusion}
\label{sec:conclusion}
In this work, we presented ARLang, a mobile augmented reality second language vocabulary learning application targeted for outdoor contexts. We evaluated ARLang (N=$44$) regarding learning effectiveness and user experience by comparing it with a flashcard app. Despite prior evidence suggesting better performance in the AR condition, our results did not show a significant difference in short-term recall between the two groups. However, participants preferred using ARLang and expressed a desire for more lessons in AR. This finding is relevant because taking learners' preferences into consideration can increase their receptiveness to extensive and complex topics \cite{ilin2022role}. Deepening our understanding of how AR experiences impact L2 learning in outdoor contexts would support the instructional design of situated learning experiences in outdoor contexts and potentially create more effective and enjoyable learning experiences for learners anytime and anywhere. As researchers find more evidence AR favors L2 learning in outdoor contexts, mobile AR experiences could extend the benefits of learning a new language to a broader audience as wide as other outdoor AR apps (e.g., Pok\'{e}mon Go \footnote{https://pokemongolive.com}). We hope this work and our findings stimulate further work to understand the challenges in designing AR applications for outdoor learning use.

\bibliographystyle{ACM-Reference-Format}
\bibliography{arlang}

\end{document}